\documentclass{ieeeaccess}
\usepackage{cite}
\usepackage{amsmath,amssymb,amsfonts}
\usepackage{algorithmic}
\usepackage{graphicx}
\usepackage{textcomp}
\usepackage{multirow}
\usepackage{balance}
\usepackage{colortbl}
\definecolor{Gray}{gray}{0.90}
\usepackage{color,soul}
\usepackage{hyperref}

\def\BibTeX{{\rm B\kern-.05em{\sc i\kern-.025em b}\kern-.08em
    T\kern-.1667em\lower.7ex\hbox{E}\kern-.125emX}}

\begin{document}
\history{Received January 14, 2021, accepted February 4, 2021, date of publication February 16, 2021, date of current version March 4, 2021.}
\doi{10.1109/ACCESS.2021.3059595}

\title{Early Detection of Myocardial Infarction in Low-Quality Echocardiography}

\author{\uppercase{Aysen Degerli}\authorrefmark{1}, \uppercase{Morteza Zabihi}\authorrefmark{1}, 
 \uppercase{Serkan Kiranyaz}\authorrefmark{2}, \IEEEmembership{Senior, IEEE}, \uppercase{Tahir Hamid}\authorrefmark{3}, \uppercase{Rashid Mazhar}\authorrefmark{3}, \uppercase{Ridha Hamila}\authorrefmark{2}, \IEEEmembership{Senior, IEEE}, and \uppercase{Moncef Gabbouj}\authorrefmark{1}, \IEEEmembership{Fellow, IEEE}}

\address[1]{Faculty of Information Technology and Communication Sciences, Tampere University, 33720 Tampere, Finland.}
\address[2]{Department of Electrical Engineering, Qatar University, Doha, Qatar.}
\address[3]{Heart Hospital, Hamad Medical Corporation, Doha, Qatar.}
\tfootnote{This work was supported by the NSF-Business Finland Center for Visual and Decision Informatics (CVDI) Advanced Machine Learning for Industrial Applications (AMaLIA) project.}

\markboth
{Aysen Degerli \headeretal: Early Detection of Myocardial Infarction in Low-Quality Echocardiography}
{Aysen Degerli \headeretal: Early Detection of Myocardial Infarction in Low-Quality Echocardiography}

\corresp{Corresponding author: Aysen Degerli (e-mail: aysen.degerli@tuni.fi).}

\begin{abstract}
Myocardial infarction (MI), or commonly known as \textit{heart attack}, is a life-threatening health problem worldwide from which 32.4 million people suffer each year. Early diagnosis and treatment of MI are crucial to prevent further heart tissue damages or death. The earliest and most reliable sign of ischemia is \textit{regional wall motion abnormality} (RWMA) of the affected part of the ventricular muscle. Echocardiography can easily, inexpensively, and non-invasively exhibit the RWMA. In this paper, we introduce a three-phase approach for early MI detection in low-quality echocardiography: 1) segmentation of the \textit{entire} left ventricle (LV) wall using a state-of-the-art deep learning model, 2) analysis of the segmented LV wall by feature engineering, and 3) early MI detection. The main contributions of this study are highly accurate segmentation of the LV wall from low-quality echocardiography, pseudo labeling approach for ground-truth formation of the unannotated LV wall, and the first public echocardiographic dataset (HMC-QU)\footnote{The benchmark HMC-QU dataset is publicly shared at the repository \href{https://www.kaggle.com/aysendegerli/hmcqu-dataset}{https://www.kaggle.com/aysendegerli/hmcqu-dataset}.} for MI detection. Furthermore, the outputs of the proposed approach can significantly help cardiologists for a better assessment of the LV wall characteristics. The proposed approach has achieved 95.72\% sensitivity and 99.58\% specificity for the LV wall segmentation, and 85.97\% sensitivity, 74.03\% specificity, and 86.85\% precision for MI detection on the HMC-QU dataset.
\end{abstract}


\titlepgskip=-15pt

\maketitle

\section{Introduction}
\PARstart{M}{yocardial} infarction (MI) is the major cause of death in the world \cite{thygesen2012third}. Solely in the United States, nearly 4 million people suffering from cardiac pain go to the emergency every year; and more than half of the accepted patients are treated in the hospitals for their recovery \cite{stillman2011assessment}. However, this process increases the expenses for the treatment and limits the medical resources needed for all patients. According to the studies of the World Health Organization (WHO), the diagnostic indicators, such as pathological results, biochemical marker values, electrocardiography (ECG) findings, and various imaging techniques are used by cardiologists to diagnose MI in patients \cite{thygesen2012third}. Nevertheless, the pathological results are not suitable for early MI detection since they can only detect dead cells in the heart muscle, which in this case, is already too late \cite{thygesen2012third}. On the other hand, biochemical marker values (cardiac enzymes) found in the human body are useful for the diagnosis \cite{thygesen2012third}, but their specificity is relatively low \cite{stillman2011assessment}. Furthermore, the electrical activity in the heart, which is measured by ECG, cannot differentiate between MI and myocardial ischemia findings \cite{thygesen2018fourth}. In addition, the interpretation of ECG for an early MI diagnosis highly depends on the experience of the medical doctor. Moreover, the ECG also relatively depicts the MI with a significant delay compared to the imaging technique so that non-diagnostic ECG still maintains as an unsolved problem \cite{sadeghi2019value}. Therefore, the most useful tool in the early diagnosis is an imaging technique, called echocardiography, which is suitable for both clinical and research purposes. The developments in echo support the diagnosis of cardiovascular diseases and its treatment as it provides an extensive assessment of the cardiovascular structure and its function \cite{gottdiener2004american, porter2018clinical}. Furthermore, echocardiography is fast, cost-effective, accessible, portable and offers the lowest risk amongst the imaging options \cite{chatzizisis2013echocardiographic, gottdiener2004american}.

Echocardiography (echo) works with the principle of ultrasound as capturing the heart muscle from different views by changing the probe angle. The human heart consists of four chambers: the right ventricle, left ventricle (LV), right atrium and left atrium in which the blood flows. The heart chambers and chamber walls can be examined in more detail with the echo devices. The focus of this paper is to assist cardiologists in diagnosing MI on the LV wall using the apical 4-chamber (A4C) view 2D echos, in which each chamber of the heart is visible. The early signs of MI are reflected as abnormalities in the chamber wall characteristics. The abnormal characteristics are ranked by \textit{hypokinesia}, \textit{akinesia} and \textit{dyskinesia} as the abnormality becomes more severe, respectively. However, these abnormalities are not easy to detect even with newer imaging technologies since the final decision is highly operator-dependent \cite{kusunose2019utilization}. Therefore, there is an urgent need for an automated, highly robust and accurate diagnostic technique that can overcome this issue.

The abnormalities of the LV wall can be captured from the LV characteristics, such as its dimension, volume, and motion via echocardiography \cite{sudarshan2014automated}. However, the evaluation process is operator-dependent, thus subjective. In order to overcome this challenge, during the last 20 years, several computer-aided techniques have been developed, which aim for more accurate and objective diagnosis \cite{doi2006diagnostic, giger2001computer}. The main techniques for MI diagnosis in echocardiography consist of active contour-based models (level sets and snake), motion estimation methods, deformation (strain) imaging, and Machine Learning algorithms. The snake approach, proposed by Kass et al. \cite{kass1988snakes}, is an elastic curve that evolves by the external constraint forces and the internal image forces in order to detect lines, boundaries, and edges in an image. It was used to track the fitted contours during motion and match them in stereopsis. Snake models have been used in several studies \cite{dong2016left, landgren2013segmentation, lin2003combinative, mishra2003ga} in echocardiography. However, when the quality of echos degrades, the snake may fail to converge to the true boundary of the LV wall, hence rendering the method useless for clinical use. 

The motion estimation algorithms are used to track pixel-based or block-based points in order to analyze the displacement of the LV wall in echo frames. Thus, the regional or global motion of the LV wall is estimated. However, the motion estimation is an ill-posed approach or even infeasible in some echos, where the noise level is high, the LV wall is not visible due to low contrast, or some part of LV wall is missing in the echo \cite{dandel2009strain, konrad1992bayesian, kordasiewicz2007affine, yu2006towards}. On the other hand, deformation imaging has become the main focus of many studies \cite{bansal2010assessment, dandel2009strain, jamal2001noninvasive, leitman2004two, nagata2015intervendor, riffel2015assessment, shah2012myocardial}. The strain is calculated from the length of the LV muscle and measured by the common method called \textit{speckle tracking}, which tracks the speckles (brightest pixels) as blocks based on a motion estimation algorithm. Consequently, the accuracy of the deformation imaging depends on the speckle tracking performance, which, once again, brings the aforementioned limitations into the strain imaging and causing different algorithms to produce unreliable results \cite{dandel2009strain}. In particular, a major limitation occurs due to temporal resolution since the minimum frame rate required for a reasonable speckle tracking is 60 frames per second (fps). 

In recent years, Machine Learning (ML) algorithms have emerged as an effective and accurate technology, which is advantageous for medical experts in solving complicated medical tasks \cite{bizopoulos2018deep}. In cardiology, many studies have been published regarding conventional and Deep Learning (DL) methods. The outstanding performance in biomedical image segmentation is achieved by the U-Net \cite{ronneberger2015u}, a supervised DL model, which is developed specifically for the segmentation tasks on the available annotated biomedical image datasets. Following its steps, many studies for the segmentation in echocardiography were published \cite{chen2016iterative, jafari2019semi, jafari2019echocardiography, leclerc2019deep, moradi2019mfp, oktay2017anatomically, smistad20172d}. However, to the best of our knowledge, there has not been any prior research to segment the \textit{entire} LV wall enclosing endocardium, myocardium and epicardium layers all at once. The segmentation of the \textit{entire} LV wall brings an advanced assessment on the LV performance parameters since the length, thickness and area of the wall segments can give valuable information related to MI. Furthermore, the number of studies to diagnose MI using the DL algorithms \cite{kusunose2020deep, omar2018automated, vidya2015computer} has increased rapidly. One major limitation is that they require large datasets for training. Even though there is a growth in the echocardiography datasets (e.g., CAMUS \cite{leclerc2019deep} and EchoNet-Dynamic \cite{ouyang2019interpretable}) that are publicly available, they are labeled for the purpose of LV volume estimation, i.e., LV ejection fraction measurement. Therefore, there is no publicly available echocardiography dataset for detection of MI.

\begin{figure*}[t!]
\centering
\includegraphics[width=1\textwidth]{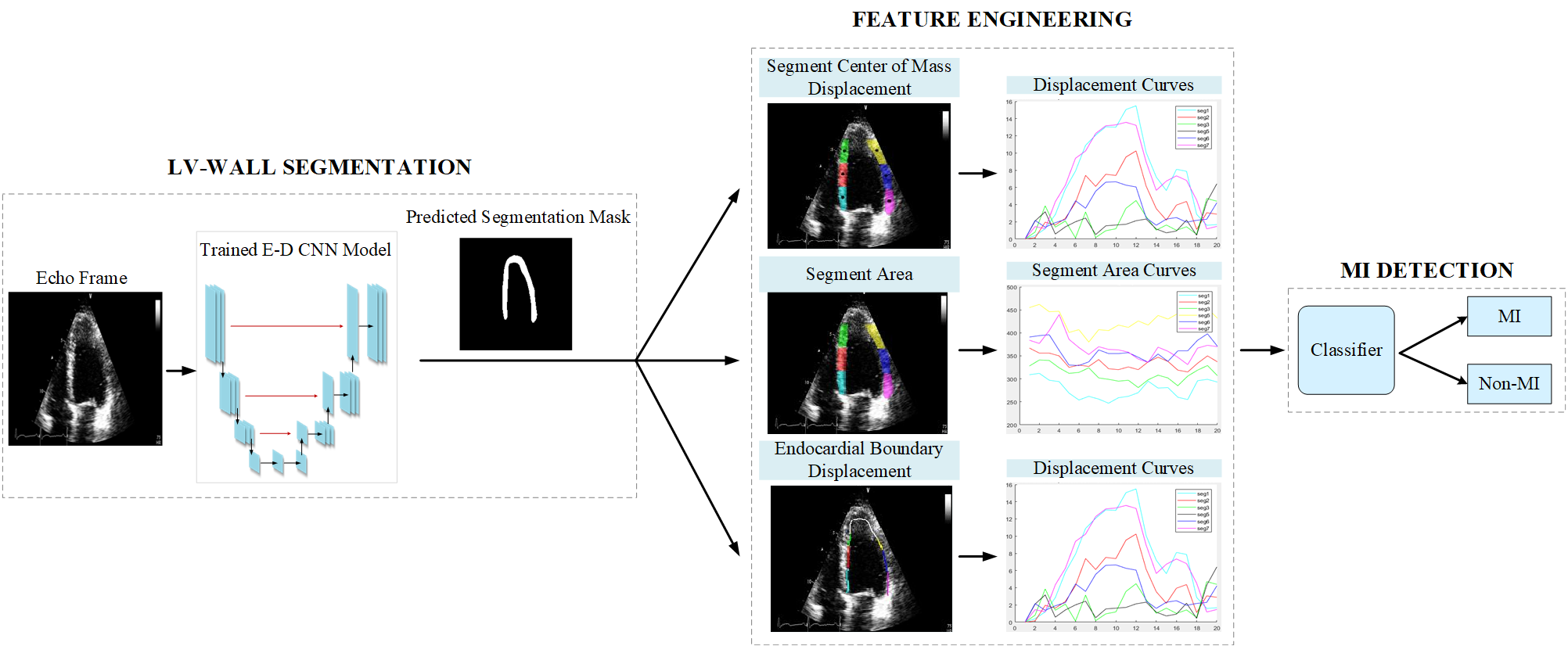}
\caption{The three-staged scheme is explained,  the first stage depicts the LV wall segmentation of each frame in an echo using the trained encoder-decoder convolutional neural network (E-D CNN) model, the second stage shows each block of feature engineering on the predicted segmentation masks, and the third block represents the MI detection by a conventional classifier.}
\label{fig:framework}
\end{figure*}

In this paper, we developed a novel three-phase approach for the early detection of MI. First, the \textit{entire} LV wall on each frame is segmented. For this purpose, the ground-truth masks for the LV wall segmentation are annotated via pseudo labeling technique (Section \ref{Pseudolabeling}). Then, the produced masks for each echo frame are verified under expert supervision that saved us numerous hours compared to manual labeling processes. After the ground-truth formation, we use an encoder-decoder Convolutional Neural Network (E-D CNN) model inspired by U-Net \cite{ronneberger2015u} (Section \ref{wallsegmentation}). Secondly, characteristics of the \textit{predicted} segmentation LV wall masks are extracted (Section \ref{characteristics}). Such characteristics include the \textit{intersections} and \textit{displacements} of LV wall segments and endocardial boundary. Finally, the features are fed into the classifiers in order to detect early signs of MI. The performance results of each classifier are compared (Section \ref{miclassresults}). Moreover, the outputs of the scheme provide advanced visualizations to cardiologists via color-coded segments and endocardial boundary illustrations on the predicted LV wall, the displacement curves of the segment center and endocardial points, and the segment area curve (Section \ref{characteristics}). As a result, cardiologists will not only have a highly accurate MI diagnosis, they will have many crucial visual cues and measurements, which in turn can help them perform a more reliable and objective assessment. 

This study further releases the 2D echocardiography dataset, HMC-QU, which is created by the cardiologists at the Hamad Medical Corporation (HMC) Hospital in Qatar along with the aforementioned bi-products generated by the proposed approach. HMC-QU is the first public dataset shared to the research community that serves both MI detection and LV wall segmentation.

\section{Materials and Methods}\label{meth}
The developed scheme consists of three stages: i) LV wall segmentation, ii) feature engineering, and iii) MI detection as illustrated in Fig. \ref{fig:framework}. In this section, we first describe the benchmark dataset, which is used in the experiments. Then, the pseudo labeling technique is introduced for the ground-truth mask formation. After the ground-truth masks are generated, the deep E-D CNN model is trained based on the masks, and the \textit{entire} LV wall is segmented in each frame of the echo. The predicted segments by the model are used to extract features based on the wall characteristics. Finally, the extracted features are fed into the ensemble of classifiers in order to perform the final MI detection in an echo. 

\begin{figure}[t!]
\centering
\includegraphics[width=0.47\textwidth]{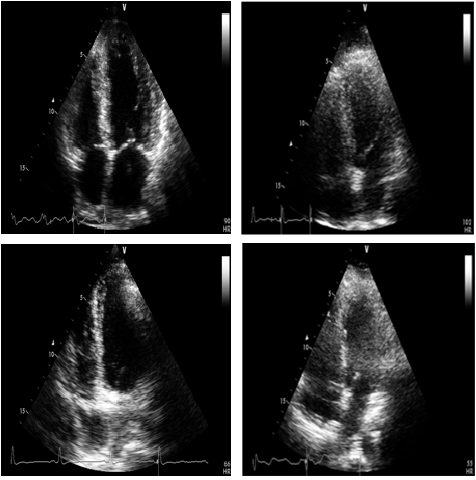}
\caption{The frames from typical low-quality echos, where LV wall is partially unrecognizable due to high level of noise with low-contrast acquisition, e.g., in the top-left sample, the upper-right section of the LV wall is entirely absent. }
\label{fig:LQ}
\end{figure}

\subsection{HMC-QU Dataset}
The HMC-QU benchmark dataset is created by the collaboration between Hamad Medical Corporation Hospital and Qatar University. The dataset includes a collection of 160 A4C echos obtained during the years 2018 and 2019. However, in this study, we used a subset of 109 echos, with a total of 2349 images from 72 MI patients and 37 non-MI subjects. The remaining 51 echos are excluded because they do not have the entire LV wall for cardiologists to evaluate (e.g., see Fig. \ref{fig:LQ}). The patients with MI were treated with coronary angioplasty after the diagnosis of acute MI with ECG and cardiac enzymes evidence. The echos from the patients are obtained before the coronary angioplasty or within 24 hours of admission to the hospital. Normal (non-MI) subjects are not diagnosed as MI but underwent a required health check in the hospital for other reasons. The usage of data has been approved by the local ethics board of HMC Hospital in February, 2019.

\begin{figure}[t!]
\centering
\includegraphics[width=0.48\textwidth]{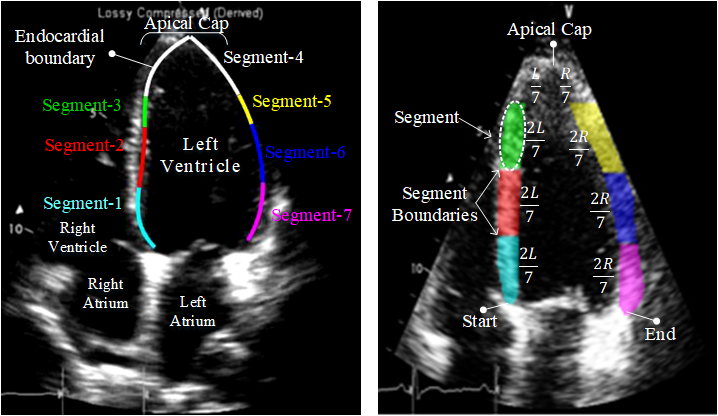}
\caption{The segment division, endocardial boundary and the heart chambers are visible on an echo frame (right side of the figure). The segment division ratios as the endocardial boundary considered to be separated into the right (from start to apical cap) and left (from apical cap to end) parts, the total length of the left part is represented as \textit{L} whereas the right part is \textit{R} (left-side of the figure).}
\label{fig:segments}
\end{figure}

The ground-truth labels of MI detection are provided unanimously by the cardiologists at HMC Hospital for the six segments (see Fig. \ref{fig:segments}) of each echo  as 1-normal, 2-hypokinesia, and 3-akinesia. For the sake of a straightforward evaluation, we have downsized the labels to two classes as 1-normal (non-MI) and 2-infarcted (MI). Table \ref{tabsegments} shows the number of the subjects: segment and patient (video) regarding MI and non-MI. There is a clear imbalance among the numbers of MI segments, which makes this problem more challenging.

The devices used for acquisition are Phillips and GE Vivid (GE-Healthcare-USA) ultrasound machines. The temporal resolution of each video is 25 frames per second and the spatial resolution varies from 422$\times$636 to 768$\times$1024 pixels. However, all the frames are resized to 224$\times$224 in order to have suitable input dimensions for many state-of-the-art deep network topologies. The echo frames are not subjected to any pre-processing such as denoising since the noise level varies fairly between the echos in the dataset. Therefore, a filter type that would be sufficient for all the echos cannot be determined. Additionally, each echo is analyzed within one cardiac cycle. End-diastole and end-systole frames are defined according to the ECG recordings of the patients. For the patients who have not any ECG recordings, the cardiac cycle is defined according to the frames, where the LV area is the largest and smallest.

\begin{table}[t!]
\centering
\caption{The number of samples in MI and non-MI patients with respect to the LV wall segments}
\resizebox{.48\textwidth}{!}{
\begin{tabular}{|c|c|c|} \hline
\rowcolor[gray]{.8}LV wall segments & \# patients with MI & \# non-MI patients\\ \hline \hline
Segment-1 & 24 & 85 \\
\rowcolor{Gray}Segment-2 & 43 & 66 \\
Segment-3 & 59 & 50 \\
\rowcolor{Gray}Segment-5 & 44 & 65 \\
Segment-6 & 25 & 84 \\
\rowcolor{Gray}Segment-7 & 15 & 94 \\ \hline \hline
Patient-based & 72 & 37 \\ \hline
\end{tabular}}
\label{tabsegments}
\end{table}

\begin{figure}[t!]
\centering
\includegraphics[width=0.48\textwidth]{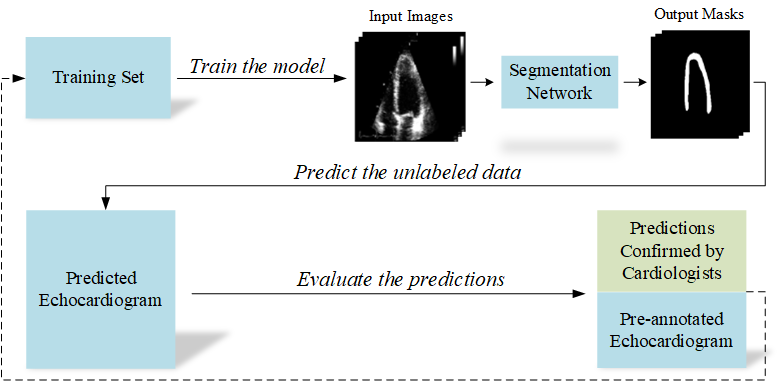}
\caption{The pseudo labeling diagram for the ground-truth annotation process of the benchmark dataset using the E-D CNN model.}
\label{fig:labeling}
\end{figure}

\subsection{Ground-truth Pseudo Labeling for Segmentation}\label{Pseudolabeling}
Deep CNN models require a large number of labeled samples for training. Expert LV wall annotation of each echo frame is cumbersome and not practical due to the number of frames in each echo. Therefore, we use a pseudo labeling technique as illustrated in Fig. \ref{fig:labeling}. First, a few ground-truth segmentation masks for the LV wall in each echo frame are provided by the cardiologists. The number of the manually segmented echos can vary among certain datasets and applications since the spatial \& temporal resolution of echo devices and patterns searched within the echo frames or videos may differ for specific applications. The echo frames, which were initially segmented by the cardiologists are used to train the E-D CNN model, which is inspired by the structure of \cite{ronneberger2015u}. Then, the trained network is used to segment the frames of other echos, which have no ground-truth segmentation masks. The initial training dataset is enriched with the correct masks that are selected by cardiologists among all the masks generated from the E-D CNN model. The enriched training set is then used to train the next E-D CNN model, and so on. Since the \textit{selection} of accurate masks visually among the model-created masks is faster (even instantaneous for a cardiologist) than manual creation, such an approach saves valuable expert labor time and costs to create a sufficiently large training set with ground-truth LV wall masks.  
 
\begin{figure*}[t!]
\centering
\includegraphics[width=0.95\textwidth]{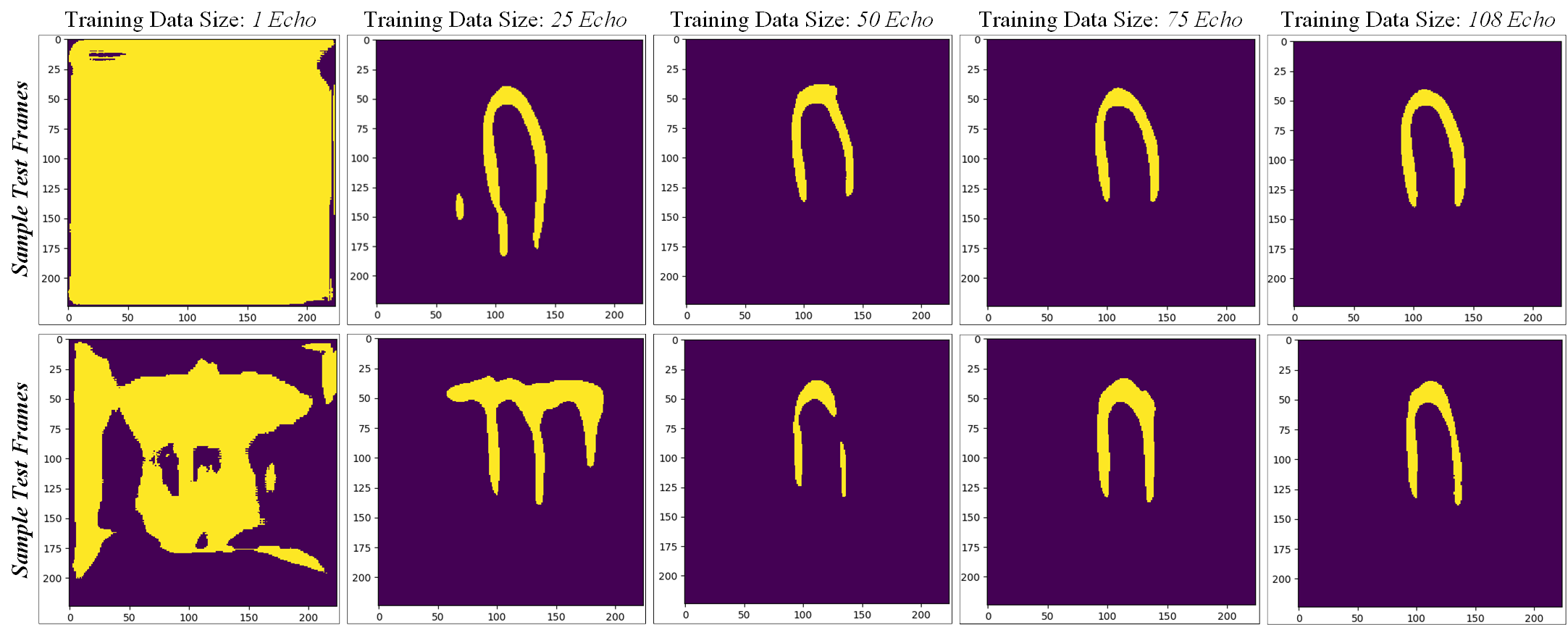}
\caption{Ground-truth pseudo labeling: The predicted segmentation masks of sample frames randomly chosen from two different echos are tested on the E-D CNN network, which is trained iteratively as the training set expands.}
\label{fig:training}
\end{figure*}

The whole process is repeated several times until the quality of the LV wall ground-truth masks are ensured. Between iterations, we perform post-processing in order to remove noise and other false positives from the model predictions while preserving the shape and size of the detected LV wall. For this purpose, we use the morphological opening operation, which is erosion followed by dilation, using a kernel with values of 1 and a size of 3x3. After each iteration, only the challenging echos remain. This gradually improves the performance of the E-D CNN model as more training data become available at each iteration as illustrated in Fig. \ref{fig:training}. Therefore, this technique can also be used for detecting noisy or problematic echos, where some parts of the LV wall are invisible. In this study, we observed that this approach could not segment 10 echos. Visual inspection confirms that these videos are either extremely noisy or some parts of the LV muscle are indeed missing (see Fig. \ref{fig:LQ}). 

\subsection{LV Wall Segmentation}\label{wallsegmentation}
The LV wall segmentation is the first step of the proposed approach as depicted in Fig. \ref{fig:framework}. Once the segmentation masks for the whole dataset are created using the pseudo labeling technique, the same network topology is used to predict LV wall in all echo frames. The model architecture is inspired by \cite{ronneberger2015u}, an encoder-decoder model, where its structure details can be seen in Table \ref{tabstructure}. The E-D CNN model is trained over the final train set, then the trained model is used to segment the LV wall of each frame of each echo in the test set. The details of the training process is given in Section \ref{segmentationresults}.  

\begin{table}[ht!]
\centering
\caption{The structure of the E-D CNN model, from encoder block through the decoder block}
\resizebox{.48\textwidth}{!}{
\begin{tabular}{c|cc|cc|} 
\cline{2-5}
 & \multicolumn{2}{c|}{Encoder Block}  & \multicolumn{2}{c|}{Decoder Block} \\ \hline \hline
\rowcolor[gray]{0.90}\multicolumn{1}{|c|}{Kernel Size} & Filters & Max Pooling & Filters & Up Sampling \\ \hline \hline
\multicolumn{1}{|c|}{3x3} & 32 & 2x2 & 512 & 2x2 \\
\rowcolor[gray]{0.95}\multicolumn{1}{|c|}{3x3} & 64 & 2x2 & 256 & 2x2 \\
\multicolumn{1}{|c|}{3x3} & 128 & 2x2 & 128 & 2x2 \\
\rowcolor[gray]{0.95}\multicolumn{1}{|c|}{3x3} & 256 & 2x2 & 64 & 2x2 \\
\multicolumn{1}{|c|}{3x3} & 512 & 2x2 & 32 & 2x2 \\
\rowcolor[gray]{0.95}\multicolumn{1}{|c|}{3x3} & 1024 & - & - & - \\ \hline
\end{tabular}}
\label{tabstructure}
\end{table}

Once the segmentation mask of the LV wall is predicted, it is divided into standardized segments as shown in Fig. \ref{fig:segments}. In this study, we have adapted the standardized model, which was recommended by the American Heart Association Writing Group on Myocardial Segmentation and Registration for Cardiac Imaging \cite{lang2005recommendations}, where the LV wall is divided into 7-segments for the A4C view. The division is done based on the endocardial boundary, which is separated into two parts as left (from the start point to the apical cap) and right (from the apical cap to the end). The length of the left part is represented as \textit{L} and the right part as \textit{R} in Fig. \ref{fig:segments}. After the segment division, the color-coded segmentation outputs (as shown in Fig. \ref{fig:segments}) are plotted as an enhanced visual evaluation for the cardiologists.

\subsection{Feature Engineering}\label{characteristics}
In this section, the segments on the LV wall are analyzed in order to capture a possible MI signature. The standardized model recommends dividing the LV wall into seven segments. However, in the analysis we only consider six of them since the apical cap, where segment-4 exists, does not exhibit inward motion activity; therefore, it should be skipped for A4C view \cite{lang2015recommendations}. For MI detection, we have extracted three different signals from the six-segments: the displacement of the endocardial boundary points, the displacement of (the center of) segments and the segment areas (see Fig. \ref{fig:framework}). In this way, we evaluate the rate of displacement from the captured global motion of the LV wall. Thus, we aim to mimic a typical diagnosis of cardiologists who assess segments that show a lack of motion as abnormal.
  
After the segmentation of the LV wall, we further extracted its inner border to define the endocardial boundary as shown in Fig. \ref{fig:framework}. Then, the boundary is divided into standardized six-segments as illustrated in Fig. \ref{fig:segments} (left). The boundary segment displacements are calculated through an echo as $L_1$ norm as follows:
\begin{equation}
    d_{L1} =  |x^{t} - x^{t_{r}}| + |y^{t} - y^{t_{r}}|
    \label{eq1}
\end{equation}
where $x$ and $y$ are the pixel coordinates of current frame $t$ and reference frame $t_{r}$ (the first frame of one cycle). In order to capture the boundary segment motion more precisely, we take $N$ times uniformly sampled pixels $p \in \{(x_1, y_1), (x_2, y_2),..., (x_N, y_N)\}$ on each frame $t$ for each segment $s$, and calculate the pair-wise distances $d_{s_t}$ between $t_{r}$ and $t$. Then, the segment displacement for each frame is calculated as in \eqref{eq2};
\begin{equation}
    d_{s_t} =  \frac{1}{N}\sum_{n=1}^{N} |x_{n}^{t_r}-x_n^t|+|y_{n}^{t_r}-y_n^t|
    \label{eq2}
\end{equation}
\begin{figure}[t!]
\centering
\includegraphics[width=0.48\textwidth]{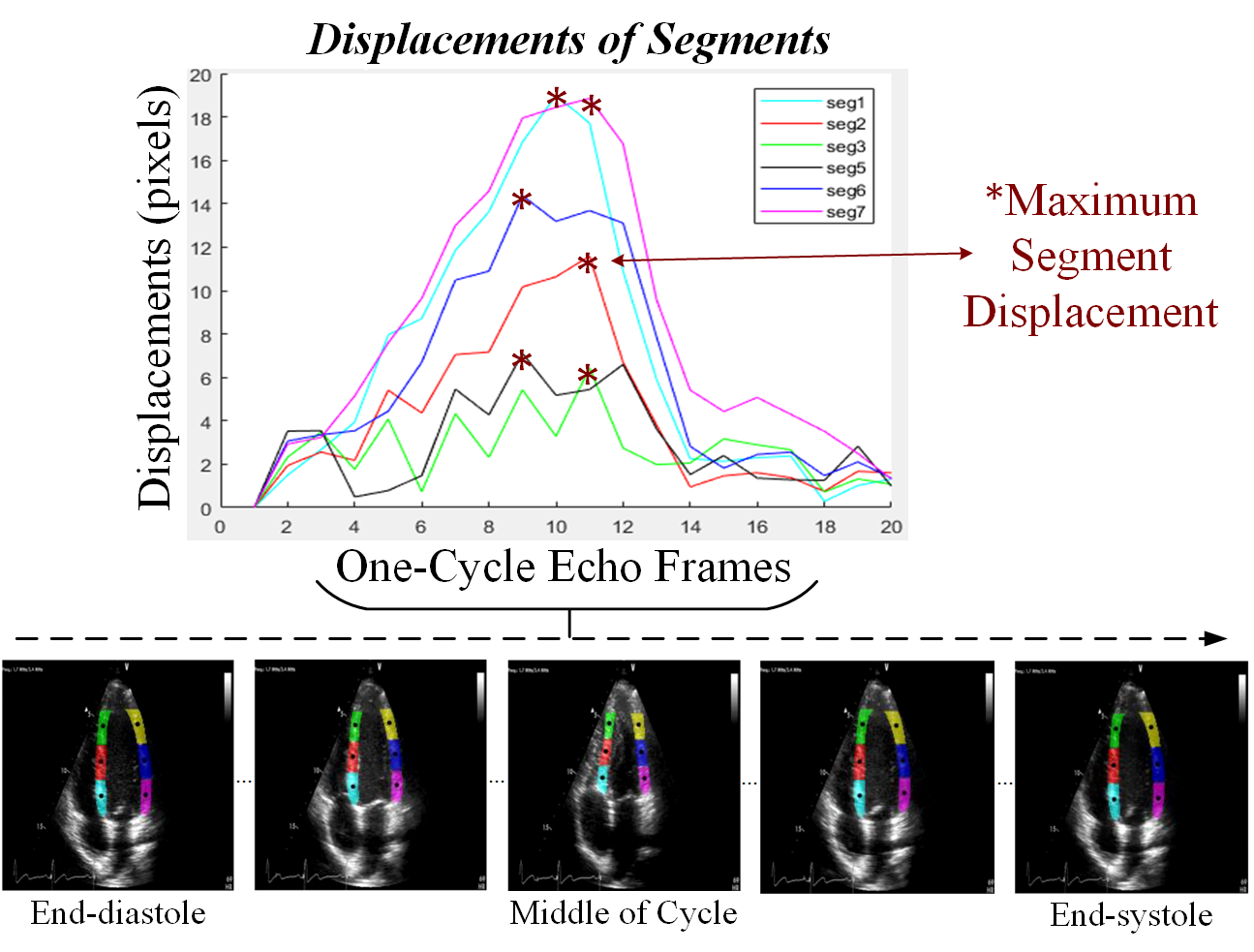}
\caption{Motion feature extraction process from end-diastole to end-systole frames in one-cycle echo. Each segment displacement curve are plotted and their maximum displacements are selected.}
\label{fig:MF}
\end{figure}
Therefore, we can obtain the displacement curve $d_s$, for each segment on the endocardial boundary of the LV wall, from which we can extract certain motion features. Moreover, the same analysis is also applied to the segment center points. This time, there is only one point on each segment to track its motion through all frames. Thus, Eq. \eqref{eq1} is still valid except that the calculation does not include the averaging step. Then, the displacement curve is plotted for the center of mass points of the six segments. Furthermore, since the \textit{entire} LV wall is extracted, we can obtain the area information from each segment. The segment area is defined as the total number of pixels included in one segment. In this way, the segment area curves can also be plotted. The overlaps within the consecutive frames' areas yield information related to segment motion and deformation since the overlapped area of the normal segments will be smaller than the ones for the infarcted segments.
  
In summary, we have extracted three sets of features: endocardial boundary motion, segment (center of mass) motion, and segment intersection area. The cardiologists visually evaluate the LV wall motion from the A4C echos by capturing the infarcted segments having an \textit{attenuated} motion compared to the others. Therefore, we define both motion features (endocardial boundary and segment center of mass) as maximum displacements of the segments in one-cycle of echo as illustrated in Fig. \ref{fig:MF}. To be more specific, we take the maximum pixel displacement of each segment, $d_s^{max} = max(d_{s_t})$ from the displacement curves, and normalize it to unity. Thereby, the motion feature, $MF$ is defined as in \eqref{eq3};
\begin{equation}
    MF={d_s^{max}}
    \label{eq3}
\end{equation}
The motion feature extractions are valid for both endocardial and segment center displacements; except their middle points are different, i.e., the middle point of the segment is the center of mass. In order to compute the area feature, first we calculate the number of pixels, $P$ inside the intersected segment areas as defined in \eqref{eq4};
\begin{figure}[t!]
\centering
\includegraphics[width=0.48\textwidth]{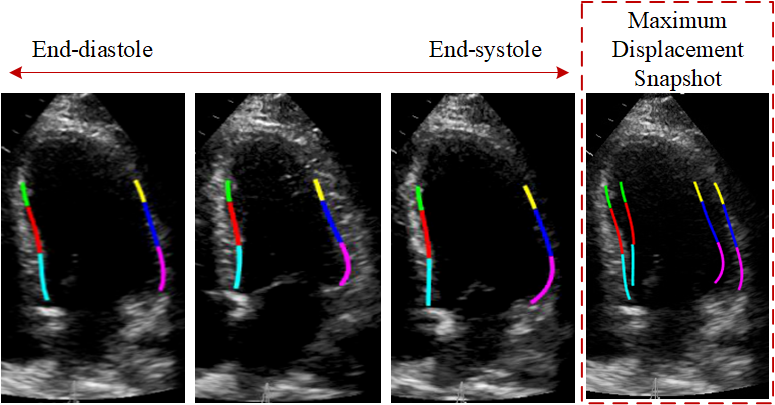}
\caption{Three frames (start-mid-end) from a sample echo, where the predicted endocardial boundary segments are color-coded from \textit{end-diastole} to \textit{systole}, and the maximum displacements of each segment is depicted.}
\label{fig:maxdisp}
\end{figure}
\begin{equation}
    P_{s_t}=P_t\cap P_{t_r}
    \label{eq4}
\end{equation}
where $P_{s_t}$ is the number of intersected pixels for segment $s$, frame $t$ calculated between the reference frame, $t_r$ and current frame, $t$. Then, the segment area feature, $AF$ is calculated as follows:
\begin{equation}
    AF=\frac{P_{s}^{min}}{P_{t_r}}
    \label{eq5}
\end{equation}
where $P_{s}^{min}$ is the minimum intersected area of a segment as $min(P_{s_t})$, and $P_{t_r}$ is the number of pixels in the reference frame segment area. Intersections can give valuable information related to MI since the larger the intersected area is, the smaller the segment movement will be in any direction. 

\begin{table}[t!]
\centering
\caption{The descriptions of the features extracted from the LV wall segments for one-cardiac-cycle echo}
\resizebox{.48\textwidth}{!}{
\begin{tabular}{|c|l}
\hline
\rowcolor[gray]{.75}\multicolumn{1}{|c|}{Features} & \multicolumn{1}{c|}{Feature Description} \\ \hline \hline
Motion Feature & \multicolumn{1}{l|}{Max displacement of endocardial boundary points} \\
\rowcolor[gray]{.9}Motion Feature &  \multicolumn{1}{l|}{Max displacement of segment center of mass points}\\
Area Feature & \multicolumn{1}{l|}{Min area intersection of segments} \\ \hline
\end{tabular}}
\label{featuredescription}
\end{table}

Table \ref{featuredescription} gives a brief description of each of the aforementioned features. We extract three features from six segments, in total of 18 features from each echo. Additionally, several crucial bi-products are created that can help cardiologists for a better and more objective assessment. These are i) color-coded LV wall and endocardial boundary segments, ii) segment (center) and endocardial boundary displacement curves, segment area curves as depicted in Fig. \ref{fig:framework}, and iii) maximum displacement snapshot of the endocardial boundary as shown in Fig. \ref{fig:maxdisp}. 

\subsection{MI Detection}
In the last stage, we used several conventional ML methods in order to detect MI in an echo. The supervised ML techniques that we used for binary classification are Linear Discriminant Analysis (LDA), Decision Tree (DT), Random Forest (RF), and Support Vector Machine (SVM). The methods analyze the extracted features by searching for any pattern or inference from the given data. LDA clusters two or more classes by maximizing the ratio between intra-class and inter-class variances to achieve maximum separability of the classes. It is an efficient classifier for imbalanced datasets, where the number of class members is unequal \cite{balakrishnama1998linear}. DT is a hierarchical structured model, which consists of branches (conjunctions), nodes (attributes) and leaves (class labels) \cite{breiman1984ra}. The tree-like structure of DT feeds the data through the branches bypassing the nodes in order to achieve the most suitable leaf to perform a classification task. Moreover, it is beneficial to use DT on small datasets. Another version of tree classifiers is RF \cite{breiman2001random}, which overcomes the overfitting problem of DT. It is an ensemble of individual trees and performs a classification task by minimizing the correlation within them. The majority voting determines the best tree as the final model, which will then be used for the rest of the task. Lastly, we used SVM classifier, an efficient classification method that clusters the data through a hyperplane \cite{cristianini2000introduction}. It is both suitable for multi-class and binary classification tasks and the kernel trick can be performed by mapping the data into a higher dimension, where it becomes easily separable. 

We have experimented with such conventional ML techniques rather than complex Deep Learning (DL) methods since our dataset is small and imbalanced for such deep models. Furthermore, DL is more suitable for complex structured data in high dimensions, whereas the extracted features lend themselves to a simpler analysis. The classifiers are evaluated in a stratified 5-fold cross-validation scheme for fair performance evaluation. Their configuration, training and testing details are explained in the next section.

\section{Experimental Results}\label{exper}
The performance evaluation of the proposed approach is carried out for both LV wall segmentation and MI detection problems. The elements of the confusion matrix are computed as follows; true negative (TN) is the number of correctly detected background pixels, true positive (TP) is the number of correctly detected LV wall pixels, false negative (FN) is the number of false detected LV wall pixels as background, and false positive (FP) is the number of false detected background pixels as LV muscle. For the MI detection, we consider the infarcted class, MI, as class-\textit{positive} and normal, non-MI as class-\textit{negative}. In this case, the confusion matrix is formed as; TN is the number of correctly predicted non-MI subjects, TP is the number of correctly predicted MI patients, FN is the number of incorrectly detected MI patients as non-MI subjects, and FP is the number of incorrectly detected non-MI subjects as MI patients. The confusion matrix elements are calculated per-frame at the pixel-level for the LV wall segmentation and per-video for the MI detection. The standard performance evaluation metrics are defined as follows:
\begin{equation}
    R=\dfrac{TP}{TP+FN}
\end{equation}
where \textit{R (recall or sensitivity)} is the ratio of correctly detected positive samples to all positive class members,
\begin{equation}
    SPE=\dfrac{TN}{TN+FP}
\end{equation}
\textit{SPE (specificity)} is the ratio of correctly detected negative samples to all negative class members,
\begin{equation}
    P=\dfrac{TP}{TP+FP}
\end{equation}
\textit{P (precision)} is the rate of correctly predicted positive class members in the all members detected as a positive class,
\begin{equation}
    F1 =\dfrac{2TP}{2TP+FP+FN}
\end{equation}
\textit{F1} is the harmonic average of \textit{precision} and \textit{sensitivity},
\begin{equation}
    ACC=\dfrac{TP+TN}{TP+TN+FP+FN}
\end{equation}
\textit{ACC (accuracy)} is the rate of all the correctly predicted classes among all the data. Accuracy might be a misleading performance metric when the dataset is imbalanced. The main objective of the analysis is to obtain the highest possible \textit{sensitivity}, with a reasonably high \textit{specificity} in order to not miss any LV wall pixels or patient with MI. 

\subsection{LV Wall Segmentation Experiments}\label{segmentationresults}
The model is evaluated by stratified 5-fold cross-validation (CV) scheme. To be more specific, we train the model using 80\% of available echos in the dataset and test it on the remaining 20\% holdout (unseen) echos. In the training process, we use Adam optimization algorithm \cite{kingma2014adam} along with a cross-entropy loss function to train the E-D CNN model with 32 mini-batch sizes, 25 epochs and a learning rate of 10$^{-3}$ in each fold. The model is implemented in Keras using Tensorflow backend on NVIDIA GeForce GTX 1080 Ti GPU.

\begin{table}[h!]
\centering
\caption{The segmentation stratified 5-fold CV folds performance results (\%) on the test (unseen) fold of the E-D CNN structure}
\resizebox{.48\textwidth}{!}{
\begin{tabular}{|c|c|c|c|c|c|}\hline
\rowcolor{Gray} CV Folds & Sensitivity & Specificity & Precision & F1 & Accuracy \\ \hline \hline
\multicolumn{1}{|c|}{Fold-1} & 94.90 & 99.70 & 93.66 & 94.26 & 99.49 \\
\multicolumn{1}{|c|}{Fold-2} & 96.53 & 99.67 & 92.31 & 94.37 & 99.54 \\
\multicolumn{1}{|c|}{Fold-3} & 96.35 & 99.63 & 91.63 & 93.93 & 99.50 \\
\multicolumn{1}{|c|}{Fold-4} & 97.61 & 99.17 & 83.75 & 90.15 & 99.10 \\
\multicolumn{1}{|c|}{Fold-5} & 93.23 & 99.75 & 94.24 & 93.73 & 99.48 \\ \hline
\rowcolor{Gray}\multicolumn{1}{|c|}{Mean} & \textbf{95.72} & \textbf{99.58} & \textbf{91.11} & \textbf{93.29} & \textbf{99.42} \\ \hline
\multicolumn{1}{|c|}{Std} & $\pm$1.69 & $\pm$0.23 & $\pm$4.25 & $\pm$1.77 & $\pm$0.18 \\ \hline
\end{tabular}}
\label{tabsegmentation}
\end{table}

\begin{figure}[t!]
\centering
\includegraphics[width=0.46\textwidth]{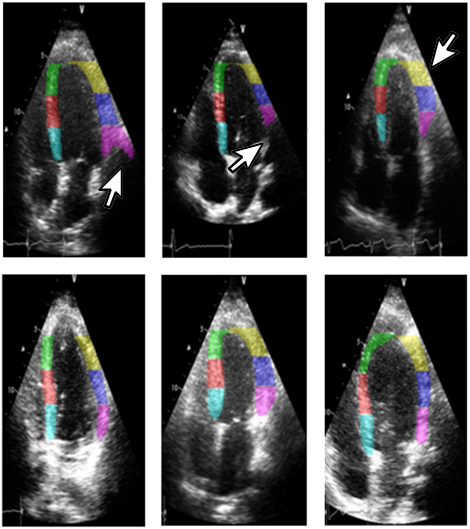}
\caption{The prediction masks of several echo frames with low-quality, which are noisy or with missing LV wall sections. The upper row shows some of the failure cases, whereas the bottom row shows successful segmentation of the LV wall on the low-quality echos.}
\label{fig:goodandbad}
\end{figure}

Table \ref{tabsegmentation} shows the LV wall segmentation results for each 5-fold CV and their averages (mean). The results indicate the robustness of the model as the proposed approach can achieve a high segmentation accuracy by 99.42\% with an F1$>$93\% on average. The segmentation evaluation is performed on a pixel-level. In fact, considering the low temporal resolution and poor quality of many videos, the E-D CNN model trained over the iterative pseudo labeled dataset achieved an elegant pixel-level performance on specificity by 99.58\% which has significantly reduced the false LV wall pixels in the segmentation stage. Additionally, 95.72\% sensitivity level on the average ensures a robust MI analysis on the next (analysis) stage. The LV wall segmentation over a low-quality echo is very challenging as it is depicted in Fig. \ref{fig:goodandbad}. Even though the results show that the proposed method is quite robust in terms of segmentation accuracy; the error can still deteriorate the following motion analysis, which will be covered in the next section. 

Previous studies on LV wall segmentation extract only the endocardial boundary of the wall. However, in this study, in addition to endocardial boundary of the LV wall, we further extract the \textit{entire} LV wall for  analysis. Thus, our segmentation method can potentially provide more information regarding wall characteristics and can exhibit high robustness against noise and artifacts.

\subsection{MI Detection Experiments}\label{miclassresults}
The performance of MI detection is evaluated on the same 5-fold CV. In each fold, the trained model is evaluated over its test batch. From each batch, we extract the aforementioned features in Section \ref{characteristics} and fed them to different classifiers: Linear Discriminant Analysis (LDA), Decision Tree (DT), Random Forest (RF) and Support Vector Machine (SVM). The models are trained by the train batch of the fold and evaluated on the (unseen) test batch. The feature engineering and MI detection stages of the scheme are implemented on MATLAB version 2019a. The SVM classifier is implemented by the Library for Support Vector Machines - LIBSVM \cite{CC01a}. The hyperparameter search is performed for each classifier to determine the optimal parameters that yield max \textit{F1-score}. The parameters are as follows; the discriminant type of LDA is set to \textit{pseudolinear} and its prior probability to \textit{uniform}, the number of tree for RF is set to \textit{10}, the pruning criterion of DT is set to \textit{impurity}, and the type of SVM is set to \textit{C-SVC} with \textit{radial basis function (rbf)} as kernel, and a cost value of \textit{10}.

Table \ref{tabmidetection} shows the results of each classifier and their performances on the features extracted from all the 6-segments. All the classifiers are evaluated by stratified 5-fold CV and the table shows the averages of the folds. The most crucial metric for MI detection is \textit{sensitivity} since the aim is to not miss any patient with MI. In fact, the SVM classifier holds the leading results with the highest sensitivity by 85.97\%, elegant specificity by 74.03\% and precision by 86.85\%. 

\begin{table}[t!]
\centering
\caption{The 5-fold CV averages of MI detection performance results (\%) of the classifiers}

\resizebox{.48\textwidth}{!}{
\begin{tabular}{|c|c|c|c|c|c|} 
\rowcolor[gray]{0.75}\multicolumn{6}{c}{\textit{6-segment Features}} \\ \hline
Classifier & Sensitivity & Specificity & Precision & F1 & Accuracy \\\hline \hline
\rowcolor{Gray}LDA & 78.51 & 70.10 & 83.89 & 80.67 & 75.65\\
               DT  & 79.09 & 58.60 & 80.41 & 79.48 & 72.62 \\
\rowcolor{Gray}RF  & 80.26 & \textbf{71.81} & \textbf{85.99} & 82.57 & 77.47 \\
               SVM & \textbf{85.97} & 70.10 & 85.52 & \textbf{85.29} & \textbf{80.24} \\ \hline
\end{tabular}}

\resizebox{.48\textwidth}{!}{
\begin{tabular}{|c|c|c|c|c|c|} 
\rowcolor[gray]{0.75}\multicolumn{6}{c}{\textit{5-segment Features}} \\ \hline
Classifier & Sensitivity & Specificity & Precision & F1 & Accuracy \\\hline \hline
\rowcolor{Gray}LDA & 81.38 & 72.32 & 86.61 & 83.21 & 78.52 \\
               DT  & 79.09 & 62.60 & 81.72 & 80.00 & 73.53 \\
\rowcolor{Gray}RF  & 82.29 & 67.37 & 84.94 & 82.65 & 76.61 \\
               SVM & \textbf{83.09} & \textbf{74.03} & \textbf{86.85} & \textbf{84.83} & \textbf{80.24} \\ \hline 
\end{tabular}}

\label{tabmidetection}
\end{table}

In an echo, the noise is usually the most severe on the apex (upper) part of the LV wall and this deteriorates the diagnosis of MI since the features extracted from the apex segments (\textit{especially} segment $5$) contains misleading information. Therefore, we have further investigated the effect of noise on the diagnosis of MI by excluding the features coming from the apex segment. Table \ref{tabmidetection} examines the performance of the classifiers as the features are excluded from segment $5$. The results show that by eliminating segment $5$ features, the highest \textit{specificity} of 74.03\%, and \textit{precision} of 86.85\% are achieved by SVM. 

\begin{figure}[t!]
\centering
\includegraphics[width=0.46\textwidth]{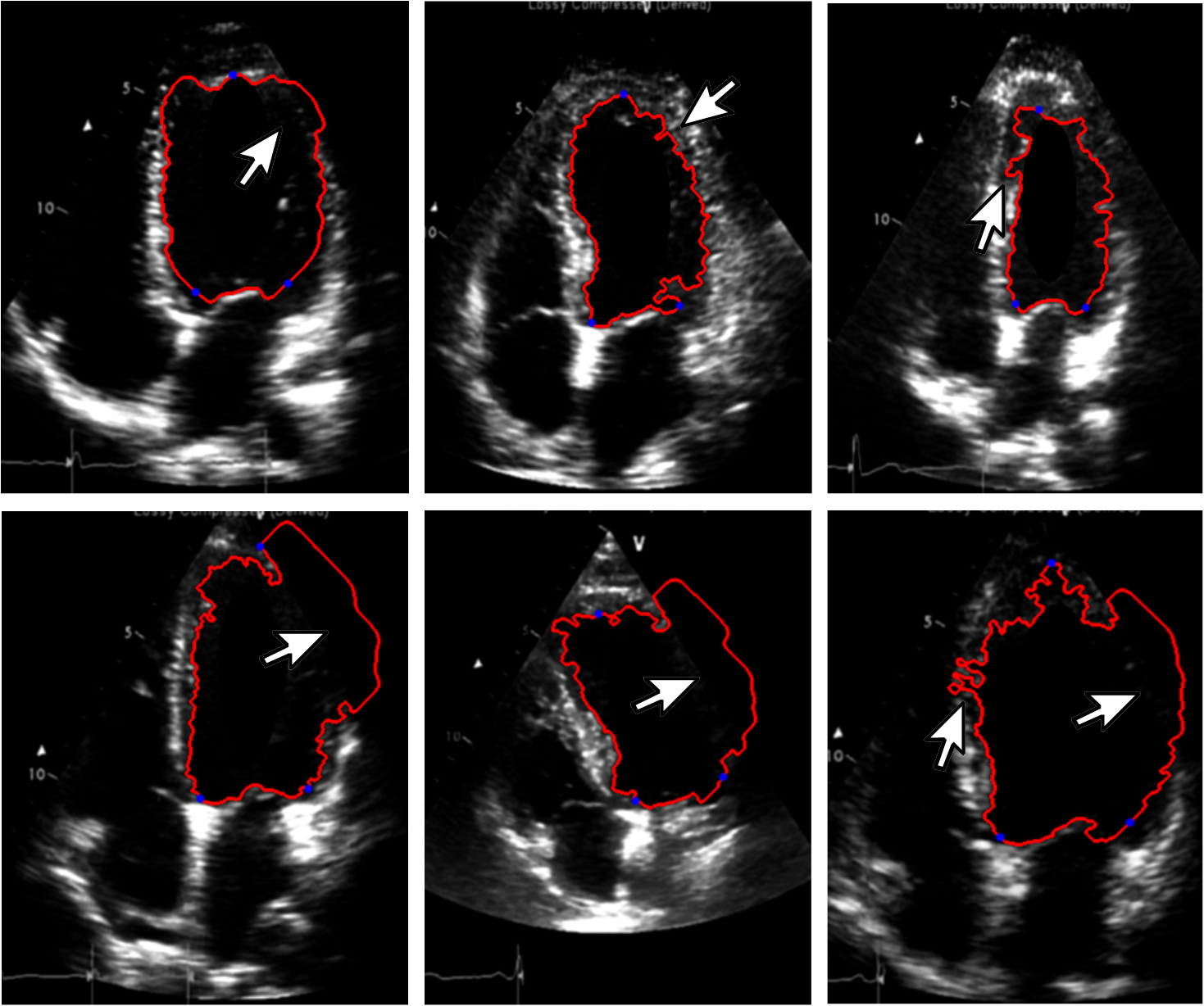}
\caption{The snake algorithm proposed by Chan-Vese \cite{chan2001active} tested on six echo frames in which it simply fails to detect the LV wall on the \textit{bottom row}, and gives reasonable but not smooth results on the \textit{top row}.}
\label{fig:snake}
\end{figure}

The direct comparison of the proposed approach against other algorithms is not possible mainly due to their inability to cope with such low-quality echocardiography. For instance, the improved version (level set formulation) of the snake-based method proposed by Chan-Vese \cite{chan2001active} simply fails badly in this dataset (see Fig. \ref{fig:snake} the bottom row). Similarly, all speckle-based approaches are not applicable to the echos in this dataset due to the poor temporal resolution, \textit{25 fps}, which is far lower than the required level \textit{60 fps} \cite{dandel2009strain}. Even if the temporal resolution would have been sufficiently high, they would still fail due to the high noise presence and especially the lack of contrast, which is visible in many echos in the dataset (see Fig. \ref{fig:LQ}). In brief, the proposed approach is so far the only feasible technique for such a low-quality echo dataset, which is in fact a commonality especially in the hospitals of many developing countries. 

\subsection{Complexity Analysis}
We have analyzed the computational complexity of our proposed approach for the three stages as follows: i) the E-D CNN segmentation model, ii) feature engineering, and iii) MI detection classifiers. The complexity of the LV wall segmentation stage for \textit{one echo frame} can be examined by the computations performed on the convolutional layers as follows;
\begin{subequations}
    \begin{equation}
        C_{mul} = \sum_{l=1}^L N_{(l-1)} N_l S_{l-1} K_{l}^2
        \label{eq:11a}
    \end{equation}
    
    \begin{equation}
     C_{add} = \sum_{l=1}^L N_{(l-1)} N_l S_{l-1} (K_{l}-1)^2 + N_{(l-1)} N_l S_{l-1}
        \label{eq:11b}
    \end{equation}
\end{subequations}
where $C_{mul}$ in \eqref{eq:11a} refers to number of the multiplication operations and $C_{add}$ in \eqref{eq:11b} refers to addition operations of $L$ layers, $N_l N_{l-1}$ is number of connections between the current and its previous layer, $S$ is the size of input feature map and $K$ is the filter length. Therefore, the time complexity of the convolutional layer is defined as follows;
\begin{equation}
O(\sum_{l=1}^L N_{(l-1)} N_l S_{l-1} K_{l}^2).    
\end{equation}

The least computationally demanding stage is the feature engineering, which consists of displacement and area calculations. The segment displacement calculation in \eqref{eq2} that approximates the pair-wise distance for $P$ number of pixels, has a complexity of $O(P)$ on average. The area intersection calculation that consists of multiplications of segment masks and summations of pixels has a complexity of $O(P)$ on average. Moreover, the maximum and minimum operations that are applied in \eqref{eq3} and \eqref{eq5} has a complexity of $O(P)$. Thus, we can define the overall complexity of the feature engineering stage for one echo as $O(Pf)$, where $P$ is number of pixels used in the calculations, and $f$ is the number of frames in an echo. 

The classifiers used in the MI detection stage for \textit{testing one echo} have the computational complexities as follows: LDA by $O(V^2)$ , DT by $O(V)$, RF by $O(Vn^{tree})$, and SVM by $O(n^{sv}V)$, where $V$ is the length of feature vector, $n^{tree}$ is the number of trees, and $n^{sv}$ is the number of support vectors.

\begin{table}[t!]
\centering
\caption{The time elapsed for executing the algorithm stages on one echo}
\resizebox{.48\textwidth}{!}{
\begin{tabular}{|c|c|c|}
\hline
\rowcolor[gray]{0.75}Algorithm Stage & Proposed Methods & Elapsed Time \textit{(ms)} \\ \hline\hline
LV Wall Segmentation & E-D CNN & 2579 \\ \hline\hline
\multirow{3}{*}{Feature Engineering} & \cellcolor[gray]{0.95}Area & \cellcolor[gray]{0.95}169.6 \\
 & Segment Center Motion & 176.9 \\
 & \cellcolor[gray]{0.95}Endocardial Motion& \cellcolor[gray]{0.95}44.9 \\ \hline\hline
\multirow{4}{*}{MI Detection} & LDA & 1.3 \\
 & \cellcolor[gray]{0.95}DT & \cellcolor[gray]{0.95}0.5 \\
 & RF & 7.0 \\
 & \cellcolor[gray]{0.95}SVM & \cellcolor[gray]{0.95}0.2 \\ \hline
\end{tabular}}
\label{tabcomputation}
\end{table}

The LV wall segmentation stage of the algorithm is implemented on a workstation with NVIDIA GeForce GTX 1080 Ti GPU and 128 GB memory, whereas the other stages are implemented on MATLAB version 2019a over a PC with 1.90 GHz CPU and 32.0 GB memory. Table \ref{tabcomputation} shows the average time elapsed for one echo (a cardiac cycle) to be executed for each operation in Fig. \ref{fig:framework}. The majority of the computational complexity originates from the LV wall segmentation stage, where ED-CNN requires 2.58 seconds to process a one-cardiac-cycle echo ($\approx20-30$ frames).

\section{Conclusions}\label{conc}
In this study, we proposed a novel three-phase approach for the early MI diagnosis from low-quality echos. For this purpose, we have created the segmentation ground-truth masks at a pixel-level for the first public 2D echocardiography dataset (HMC-QU) using the pseudo labeling technique. Then, we have used the deep E-D CNN model to segment the LV wall on each frame of each echo in the dataset. Finally, over the \textit{predicted} LV wall segments, we have extracted features and used them in several classifiers to compare their performances for MI detection. The experimental results on the HMC-QU dataset show that the developed scheme achieved an elegant performance yielding a small false alarm rate for LV wall segmentation. The achieved results by 5- fold CV for MI detection are also quite promising considering the poor quality and resolution of the echos; however, there is still room for improvement. 

The developed scheme aims for an objective and operator independent assessment by providing quantitative measurements for the LV wall motion and segment areas. Cardiologists diagnose patients with prior knowledge, such as medical history, gender, and age. Additionally, they look at the other echo views; therefore, they can interpret the movement and functionality of the heart in more detail. Hence, the performance of the developed scheme can further be improved, e.g., if the proposed features are fused together from other views, such as apical 2-chamber (A2C) and the three circular views. 

The proposed features are explicitly comprehensive since the features extracted from the segmented wall are mimicking the way that a medical expert interprets the echos. Therefore, the proposed features are valuable not only from the engineering point of view but also from a medical perspective. Additionally, the distinct visual outputs of the method, i.e., color-coded segments on the LV wall illustration, segment and endocardial boundary points displacement curve, and segment area curve plots can be crucial to the cardiologists for a better and objective assessment. The limitation of this study is that the proposed approach consists of three sequential stages for MI detection. In the future, our focus will be developing an end-to-end system to detect MI. Finally, the HMC-QU dataset along with ground-truth masks and segment labels is publicly available for the research community. HMC-QU is the first benchmark dataset that is publicly available for the research community compared to the relevant studies in this area, and it is also the largest collection ever compiled with both normal echos and echos of both male and female acute MI patients at different ages. 

\bibliographystyle{IEEEtran}
\bibliography{refs}

\newpage 

\begin{IEEEbiography}[{\includegraphics[width=1in,height=1.25in,clip,keepaspectratio]{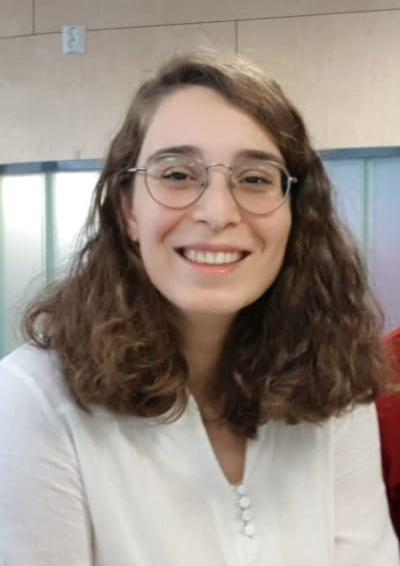}}]{Aysen Degerli} received the B.Sc. degree (Hons.) in electrical and electronics engineering from Izmir University of Economics, Turkey in 2017, and the M.Sc. degree (Hons.) in data engineering and machine learning from Tampere University, Finland in 2019. She is currently pursuing the Ph.D. degree in computing and electrical engineering with Signal Analysis and Machine Intelligence research group led by Prof. M. Gabbouj at Tampere University. Her research interests include machine learning, compressive sensing, and biomedical image processing. 
\end{IEEEbiography}

\begin{IEEEbiography}[{\includegraphics[width=1in,height=1.25in,clip,keepaspectratio]{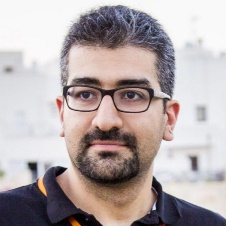}}]{Morteza Zabihi} was born in Iran, in 1988. He received the M.Sc. degree in biomedical engineering from Tampere University, Tampere, Finland, in 2013, and the Ph.D. degree from the Department of Computing Sciences, Tampere University. He is also working as a researcher with the Department of Computing Sciences. His research interests include nonlinear dynamics and time series analysis, pattern recognition, and machine learning with application to EEG and ECG signal processing. Since 2015, he has been ranked in the top three teams in four international competitions, such as PhysioNet/Computing in Cardiology and IEEE EMBS Neural Engineering Brain-Computer Interface Challenges.
\end{IEEEbiography}

\begin{IEEEbiography}[{\includegraphics[width=1in,height=1.25in,clip,keepaspectratio]{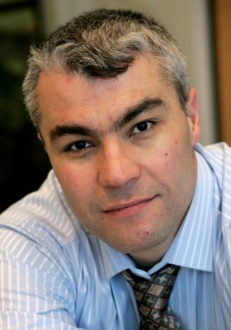}}]{Serkan Kiranyaz} is currently a Professor in Qatar University, Doha, Qatar. He published two books, five book chapters, more than 70 journal papers in high impact journals, and 100 papers in international conferences. He made contributions on evolutionary optimization, machine learning, bio-signal analysis, computer vision with applications to recognition, classification, and signal processing. He has co-authored the articles which have nominated or received the “Best Paper Award” in ICIP 2013, ICPR 2014, ICIP 2015 and IEEE TSP 2018. He had the most-popular articles in the years 2010 and 2016, and most-cited article in 2018 in IEEE Transactions on Biomedical Engineering. During 2010-2015 he authored the most-cited article of the Neural Networks journal. His research team has won the 2nd and 1st places in PhysioNet Grand Challenges 2016 and 2017, among 48 and 75 international teams, respectively. In 2019, he won the “Research Excellence Award” and “Merit Award” of Qatar University. His theoretical contributions to advance the current state of the art in modelling and representation, targeting high long-term impact, while algorithmic, system level design and implementation issues target medium and long-term challenges for the next five to ten years. He in particular aims at investigating scientific questions and inventing cutting edge solutions in “personalized biomedicine” which is in one of the most dynamic areas where science combines with technology to produce efficient signal and information processing systems meeting the high expectation of the users. 
\end{IEEEbiography}

\begin{IEEEbiography}[{\includegraphics[width=1in,height=1.25in,clip,keepaspectratio]{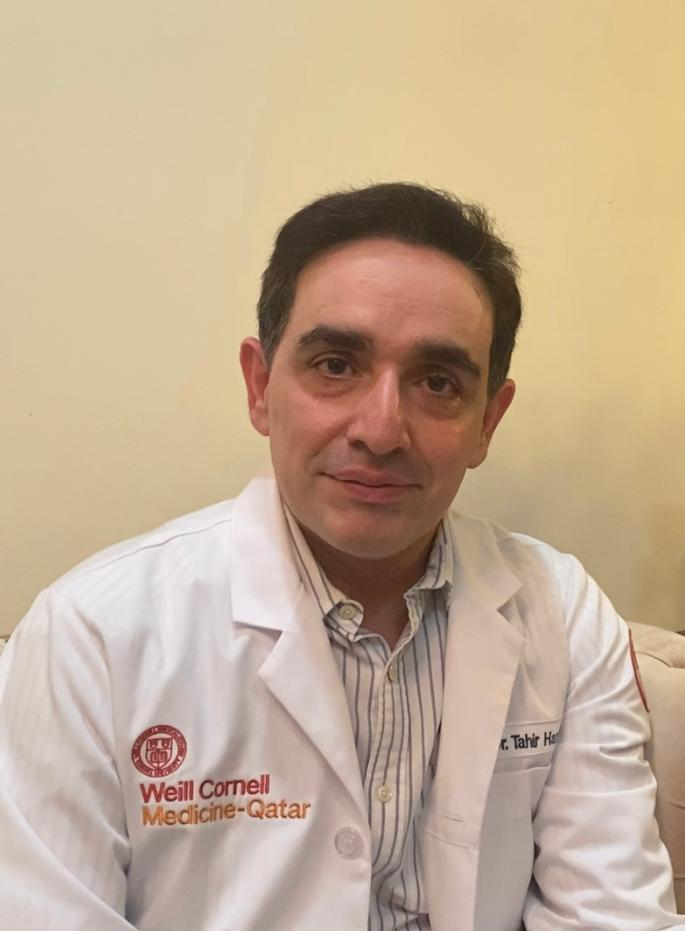}}]{Tahir Hamid} has been working as Interventional Cardiology Consultant with the Heart Hospital, Hamad Medical Corporation, Doha Qatar. He has been trained in United Kingdom and with the University of Toronto, Canada. He has about 27 publications and presented at various cardiology conferences. Currently, conducting randomized control trial about “pre-procedural fasting in patients undergoing coronary interventions” at the Toronto General Hospital, University of Toronto. He has published the following study related to ECG monitoring in patients with recurrent blackouts:  “Prolonged implantable electrocardiographic monitoring indicates a high rate of misdiagnosis of epilepsy - REVISE study”, Europace. This study revolutionized the assessment of patient blackout who were wrongly labeled as epileptics and were later found after prolonged monitoring, to have cardiac rhythm related issues leading to blackouts. He is also actively involved in the research projects at Hamad Medical Corporation. He is also working on his project about external cardiopulmonary resuscitation devices for patients who sustain cardiac arrests. 
\end{IEEEbiography}

\begin{IEEEbiography}[{\includegraphics[width=1in,height=1.25in,clip,keepaspectratio]{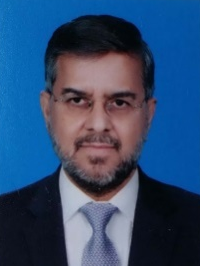}}]{Rashid Mazhar} has more than 20 years of experience in the field of cardiothoracic surgery. He is currently a Senior Consultant Thoracic surgeon with Hamad Medical Corporation, Doha, Qatar, with minimal invasive \& Robotic surgery as his areas of surgical interest. Besides his surgical, clinical and educational activities, he has a particular interest and track record of translational research, collaborating with engineers and academicians to bring about user-end driven innovative medical solutions. Automation, objective monitoring, intensive care, cardio-pulmonary resuscitation and user-friendly signal processing are his areas of interest. His research projects at Hamad Medical Corporation (HMC) and Qatar foundation has obtained grant funds in excess of 5 Million Qatari riyals.  These include five HMC, two UREP and three NPRP funded projects.  He is currently part of two NPRP funded projects. To his credit he has more than 30 publications, including a book chapter; four granted patents, four patents under process, five HMC innovations award, and four Stars of excellence awards in research category.
\end{IEEEbiography}

\newpage 
\begin{IEEEbiography}[{\includegraphics[width=1in,height=1.25in,clip,keepaspectratio]{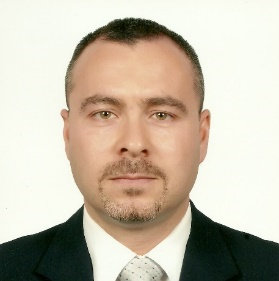}}]{Ridha Hamila} received the M.Sc., Lic.Tech. (Hons.), and Ph.D. degrees from the Tampere University of Technology (TUT), Tampere, Finland, in 1996, 1999, and 2002, respectively. From 1994 to 2002, he held various research and teaching positions at TUT within the Department of Information Technology, Finland. From 2002 to 2003, he was a System Specialist at Nokia research Center and Nokia Networks, Helsinki. From 2004 to 2009, he was with Emirates Telecommunications Corporation, UAE. Also, from 2004 to 2013, he was an Adjunct Professor at the Department of Communications Engineering, TUT. He is currently a Full Professor with the Department of Electrical Engineering, Qatar University, Doha, Qatar. His current research interests include mobile and broadband wireless communication systems, mobile edge computing, internet of everything, and machine learning. In these areas, he has published over 200 journal and conference papers most of them in the peered reviewed IEEE publications, filed seven U.S. patents, and wrote numerous confidential industrial research reports. He has been involved in several past and current industrial projects, Ooreedo, Qatar National Research Fund, Finnish Academy projects, EU research and education programs. He supervised a large number of under/graduate students and Postdoctoral Fellows. He organized many international workshops and conferences. 
\end{IEEEbiography}

\begin{IEEEbiography}[{\includegraphics[width=1in,height=1.25in,clip,keepaspectratio]{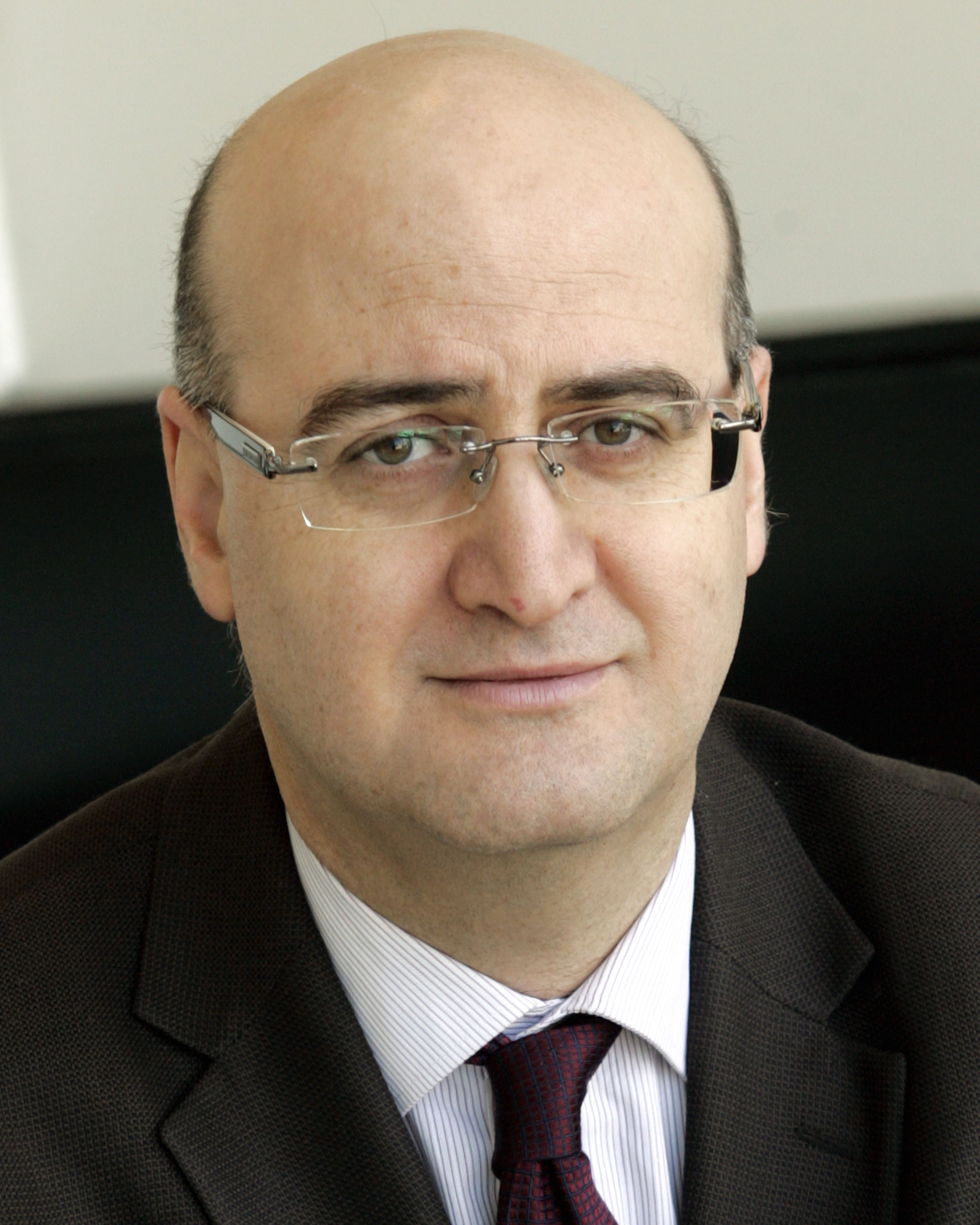}}]{Moncef Gabbouj} is currently a well-established world expert in the field of image processing, and held the prestigious post of Academy of Finland Professor from 2011 to 2015. He has been leading the Multimedia Research Group for nearly 25 years and managed successfully a large number of projects in excess of 18M Euro. He has supervised 45 Ph.D. theses and over 50 M.Sc. theses. He is the author of several books and over 700 articles. His research interests include big data analytics, multimedia content-based analysis, indexing and retrieval, artificial intelligence, machine learning, pattern recognition, nonlinear signal and image processing and analysis, voice conversion, and video processing and coding. He is also a member of the Academia Europaea and the Finnish Academy of Science and Letters. He is also the past Chairman of the IEEE CAS TC on DSP and Committee member of the IEEE Fourier Award for Signal Processing. He served as Associate Editor and Guest Editor of many IEEE, and international journals.
\end{IEEEbiography}

\EOD

\end{document}